\documentclass[aapm,mph,preprint,graphicx]{revtex4-1}

\usepackage{graphicx}
\usepackage{amsmath}
\usepackage{supertabular} 
\usepackage{fancyhdr,afterpage}
\usepackage[mathlines]{lineno}
\modulolinenumbers[5]
\linenumbers\relax 

\begin{document}%

\title{An optimized ultrasound detector for photoacoustic breast tomography}
%
%
%

\author{Wenfeng~Xia$^{a}$, Daniele~Piras$^{a}$, Johan C. G. van Hespen$^{a}$, Spiridon~van~Veldhoven$^{b}$, Christian~Prins$^{b}$, Ton~G.~van~Leeuwen$^{a,c}$, Wiendelt~Steenbergen$^{a}$, and~Srirang~Manohar$^{a,}$} 

\email{s.manohar{\it @}tnw.utwente.nl}

\affiliation{$^{a}$Biomedical Photonic Imaging group, Mira Institute for Biomedical Technology and Technical Medicine, University of Twente, P.O. Box 217, 7500AE Enschede, The Netherlands.}
 
\affiliation{$^{b}$Oldelft Ultrasound B. V., P. O. Box 5082, 2600 GB Delft, The Netherlands.}

\affiliation{$^{c}$Biomedical Engineering and Physics, Academic Medical Center, University of Amsterdam, P.O. Box 2270, 1100 DE Amsterdam, The Netherlands.}

\date{\today}

\begin{abstract}

\textbf{Purpose:}  Photoacoustic imaging has proven to be able to detect vascularization-driven optical absorption contrast associated with tumors. In order to detect breast tumors located a few centimeter deep in tissue, a sensitive ultrasound detector is of crucial importance for photoacoustic mammography. Further, because the expected photoacoustic frequency bandwidth (a few MHz to tens of kHz) is inversely proportional to the dimensions of light absorbing structures (0.5 to 10+ mm), proper choices of materials and their geometries and proper considerations in design have to be made to implement optimal photoacoustic detectors. In this study, we design and evaluate a specialized ultrasound detector for photoacoustic mammography.

\textbf{Methods:} Based on the required detector sensitivity and its frequency response, a selection of active material and matching layers and their geometries is made leading to a functional detector models. By iteration between simulation of detector performances, fabrication and experimental characterization of functional models an optimized implementation is made and evaluated. For computer simulation, we use 1D Krimholtz-Leedom-Matthaei (KLM) and 3D finite-element (FEM) based models. 

\textbf{Results:} The experimental results of the designed first and second functional detectors matched with the simulations. In subsequent bare piezoelectric samples the effect of lateral resonances was addressed and their influence minimized by sub-dicing the samples. Consequently, using simulations, a final optimized detector was designed, with a center frequency of 1 MHz and a -6 dB bandwidth of 0.4-1.25 MHz (fractional bandwidth of $\sim$80$\%$). The minimum detectable pressure was measured to be 0.5 Pa.
 
\textbf{Conclusion:}  A single-element, large-aperture, sensitive and broadband detector is designed and developed for photoacoustic tomography of the breast. The detector should be capable of detecting vascularized tumors with 1-2 mm resolution. The minimum detectable pressure is 0.5 Pa, which will facilitate deeper imaging compared to the current systems. Further improvements by proper electrical grounding and shielding and implementation of this design into an arrayed detector will pave the way for clinical applications of photoacoustic mammography. 

\end{abstract}

\maketitle


\section{Introduction}

Photoacoustic imaging is an intrinsically hybrid biomedical imaging modality which is based on light excitation and ultrasound detection. In photoacoustics, short-pulsed laser light diffusively penetrates into tissue, is selectively absorbed by specific chromophores, such as hemoglobin, causing ultrasound waves to be generated by thermoelastic expansion. These are then detected by ultrasound detectors at the surface of tissue. Since ultrasound scattering is considerably lower than light scattering in biological tissue, the information carried by ultrasound waves arrives with less losses and distortion at the tissue surface than light waves would do. Thus photoacoustic imaging takes the advantages of both optical imaging and ultrasound imaging, with a high optical contrast, a high ultrasound resolution and large imaging depths~\cite{Razansky2009, Wang2012, Wang2003, Jose2009, Beard2011, Zhu2009}. One of the most important applications of photoacoustics is breast imaging, where the optical contrast comes from higher average hemoglobin levels associated with malignant masses compared to healthy breast tissue~\cite{Tromberg2000, Tromberg2008}. The technique promises to be an alternative to ionizing x-ray and low contrast ultrasound imaging to detect breast cancer~\cite{Piras2010, Kruger2010, Ermilov2009, Pramanik2008, Kitai2012, Heijblom2012}. 

The faithful and sensitive detection of ultrasound (US) lies at the heart of a photoacoustic imaging system. The US detector largely determines image contrast, resolution and imaging depth of the system. In photoacoustic (PA) imaging of the breast, there is a requirement to detect tumors located a few centimeter deep in tissue, where light is heavily attenuated. Thus a sensitive ultrasound detector is of crucial importance. Further, the frequency range of photoacoustic waves is inversely proportional to the dimensions of absorbing structures~\cite{Khan1995}. In breast tissue, structures of interest range from 0.5 to 10+ mm initiating ultrasound frequencies from a few MHz down to tens of kHz~\cite{Allen2012}. Thus a broadband US detector is required, centered on an optimum frequency. 

Various US detectors with different specifications have been employed in photoacoustic (thermoacoustic) systems for breast imaging~\cite{Kruger2010, Pramanik2008, Kruger2000, Ermilov2009, Khan1995, Khokhlova2007, Andreev2003, Manohar2005}. In 2000, Kruger {\it et al.}~\cite{Kruger2000}, reported the first thermoacoustic (TA) breast scanner in CT configuration. The US detector used 64 commercially available immersion detectors (1-3 piezocomposite, model 3847, Panametrics, Waltham, Mass.) each with flat face and a diameter of  13 mm, arrayed in a spiral pattern spanning the surface of a hemispherical bowl. The detector had a center frequency of 1 MHz with a fractional frequency bandwidth (FBW) of 70$\%$. The sensitivity of the detector was not reported. A later version of the system from 2010~\cite{Kruger2010}, employed an array of 128 detectors (1-3 piezocomposite), 3 mm in diameter with a 5 MHz center frequency and a FBW of 70$\%$.  Submillimeter breast vasculature down to a depth of 40 mm was successfully visualized~\cite{Kruger2010}. The sensitivity was not reported. However, the authors recommended the use of detectors with a lower center frequency to have greater imaging sensitivity. 

Pramanik {\it et al.}~\cite{Pramanik2008} in 2008, reported a breast cancer detection system combining TA and PA tomography. They used 13-mm/6-mm-diameter active area non-focused detectors operating at 2.25 MHz center frequency (piezocomposite, ISS 2.25 x 0.5 COM, Krautkramer) for signal detection. The detectors used have a large active surface area and a relatively low center frequency to gain a high sensitivity. The FBW is reported as varing between 60-120$\%$ but no details of sensitivity are reported.  

The laser optoacoustic imaging system (LOIS) was developed over a decade ago and has undergone several iterations~\cite{Ermilov2009, Khokhlova2007, Andreev2003}. The latest system uses an array of 64 wideband polyvinylidene fluoride (PVDF) (bandwidth upto 2.5 MHz) elements arranged in a concave arc. Each detector element has a rectangular surface, with a large aspect ratio (20 mm x 3 mm). This design gives the detector a slice-shape focusing area, providing high sensitivity in the region of interest. The sensitivity of this system was reported to be 1.66 mV/Pa at 1.5 MHz, the minimum detectable pressure (MDP) was not reported~\cite{Ermilov2009}. However, for some earlier versions the MDP values have been presented including measured~\cite{Lamela2011} and estimated using only the thermal noise generated by the detector capacitance according to the Nyquist law~\cite{Andreev2003, Andreev2000}.  

The Twente Photoacoustic Mammoscope (PAM) has been previously developed in our group in 2004~\cite{Manohar2004}. US detection uses a planar array of 590 PVDF detector elements. Each element has a 2 mm x 2 mm active surface area. The detector has a center frequency of 1 MHz and a fractional bandwidth of 130$\%$~\cite{Manohar2005}. The measured MPD value is 80 Pa~\cite{Piras2010}. Promising clinical measurement results have been reported in 2007 and 2012~\cite{Manohar2007, Heijblom2012}. However, the system still suffers from the relatively low sensitivity of the PVDF detector.
  
PVDF detector arrays give LOIS and PAM broad bandwidths, while the measured sensitivity of the detectors are generally low compared to PZT detectors due to the reduced electromechanical coupling coefficients~\cite{Hunt1983}. However, PZT detector suffers from a relatively low bandwidth. As a compromise, piezocomposite detectors provide better sensitivity and reasonably good FBW as employed by Kruger {\it et al.}~\cite{Kruger2010} and Pramanik {\it et al.}~\cite{Pramanik2008}. 

To achieve a high sensitivity of the detector, we choose a highly sensitive PZT material, while tailoring the bandwidth of the detector to be reasonably wide.  A single element PZT detector structurally consists of the active piezoelectric material, front- and back-matching layers and a backing layer. To have both high sensitivity and broad bandwidth, the materials, their acoustic characteristics and their dimensions should be carefully chosen. Furthermore, the aperture of the detector should be optimized as there is a trade-off between the lateral resolution and sensitivity of the system~\cite{Xu2003}. 

In this paper, we present the design considerations for this specialized ultrasound detector for PA breast imaging.  We specify the most important detector output characteristics such as sensitivity and frequency response, and justify the selection of active material and matching layers and their geometries. We iterate between simulation of detector performance, fabrication and experimental characterization of functional models to arrive at an optimized implementation. For computer simulation, we use 1D Krimholtz-Leedom-Matthaei (KLM) models and 3D finite-element (FEM) based models.

\section{Design parameters}

\subsection{Sensitivity and acceptance angle}

In photoacoustic tomography, the breast is illuminated with short-pulsed laser light and the ultrasound detector scans the object ideally through 360$^{\texttt{o}}$. A requirement in this geometry is that each US detectors's acceptance angle is wide enough to detect photoacoustic signals generated throughout the entire object for each detector angular position around the object (Figure~\ref{fig:CT}). During a backprojection style reconstruction, the coherent signals from all detector angular positions are summed up to form an image of the photoacoustic sources~\cite{Hoelen1998}. For a single element square surface detector, the -6 dB acceptance angle ($\phi$) depends on the width ($W$) and center frequency ($f_{0}$) of the detector and can be expressed as~\cite{Woodward1992}:

\begin{equation}\label{eq1}
\phi = 2~\texttt{sin$^{-1}$}(\displaystyle \frac{3.79}{\mathrm{kW}})
\end{equation}

\noindent with $k$ the wavenumber. Equation 1 shows that the acceptance angle decreases as the aperture increases. However a larger surface area is important since for the same piezo element thickness, a lower MDP is attained. This is due to a lower thermal-induced noise due to the higher electrical capacitance associated with larger detectors. To use a sensitive large aperture detector without compromising the acceptance angle, Li {\it et al}~\cite{Li2008} and Pramanik {\it et al}~\cite{Pramanik2009} introduced the concept of a negative acoustic lens. Here an acoustic lens enlarges the acceptance angle of a large aperture detector improving the lateral resolution of the PA tomographic system. This provides a solution where one can use a large area detector while maintaining a wide acceptance angle.

Based on the above considerations, the choice is made for a large aperture detector with a square surface (5 mm x 5 mm) with an appropriate lens~\cite{Xia2011}, which will be discussed in a future article. The square shape is chosen for convenience in array development using traditional dicing. 

\subsection{Center frequency}
Photoacoustic breast imaging is based on the tissue optical contrast due to vascularization. This is enhanced around a tumor due to the process of angiogenesis~\cite{Folkman2000}. This process is reported to go through two phases separated by the ``angiogenic switch''. Exponential tumor growth ensues in the second phase (vascular phase), which occurs from tumor sizes of 1-2 mm in diameter~\cite{Bergers2003}. This indicates that the resolution of our system is preferably to be smaller than 2 mm. 

The upper and lower limit of the frequency range of the detector ($f_{max}$ and $f_{min}$(MHz)) can be estimated knowing the smallest and largest sphere ($a_{min}$ and $a_{max}$ (mm in diameter)) that are required to be resolved by the system using~\cite{Andreev2003}:

\begin{equation}\label{eq2}
	f_{max} = \displaystyle \frac{3\nu}{a_{min}} 
\&  f_{min} = \displaystyle \frac{0.32\nu}{a_{max}}	
\end{equation}

\noindent in which $\nu$ (mm $\mu$s$^{-1}$) is the acoustic velocity of the medium. Therefore, faithful registration of spherical tumors with diameters from 2 mm to 10 mm requires an ultrasound detector with frequency bandwidth from 2.25 MHz down to 48 kHz. To achieve the required bandwidth, the center frequency of the detector is designed to be 1 MHz, while optimizing the bandwidth of the detector to be reasonably broad.

\section{Materials and fabricated models}

\subsection{Materials}

As mentioned earlier, we prefer PZT detectors due to the superior dielectric constant, lower dielectric loss and higher coupling coefficients~\cite{Hunt1983}. A high sensitivity commercial piezoelectrical material CTS 3203HD (CTS Communications Components, Inc., Albuquerque, NM) is used for the active material due to the higher coupling coefficients compared to other common used commercial PZT materials~\cite{vanNeer2010}. A front matching layer is used to improve the sensitivity and bandwidth of the detector, and a back matching layer is used to improve the transmission of ultrasound into the backing layer and thus improve the bandwidth of the detector. The theoretical values for the suitable acoustic impedance of the matching layers ($Z_{m}$) can be determined by~\cite{Kossoff1966, McKeighen}:

\begin{equation}\label{eq3}
Z_{m} = \sqrt{Z_{w}Z_{p}}
\end{equation}

\noindent where $Z_{w}$ and $Z_{p}$ are the acoustic impedance of water/tissue (or backing) and ferroelectric ceramic, respectively. The material properties of the piezoelectric material, the front matching layer, back matching layer and the backing are carefully chosen based on their acoustic impedances, and all important acoustic properties for these materials are listed in Table~\ref{table:material}.

\subsection{Functional and test models}

\subsubsection{First functional model}

To have a rough assessment of the required characteristics of the detector such as center frequency (around 1 MHz) and bandwidth ($\geq$80$\%$), the thicknesses of the active material, front and back matching layer and backing are determined using the KLM model~\cite{Leedom1971}(See Sec. IV). The first functional model is then manufactured as shown in Figure~\ref{fig:singleelement}(a), with dimensions of each layer listed in Table~\ref{table:firstmodel}.

\subsubsection{Test models for minimizing radial resonances}

The large lateral dimensions of the detector result in radial resonances that interfere with the thickness resonance. 
To reduce the radial resonance of the element, the detector requires to be sub-diced into smaller units, which are acoustically isolated by air kerfs but electrically grouped by common electrodes to form a single composite element~\cite{Qi2000}(Figure~\ref{fig:singleelement}(b)). To identify optimum lateral dimensions of the subdiced unit, five square-shaped, bare ceramics samples (CTS 3203HD) all having a thickness of 1.625 mm were manufactured as shown in Figure~\ref{fig:singleelement}(c). The different sizes represent the possible subdicing sizes applied to the 5 mm x 5 mm element to suppress the radial resonances. 3D finite-element method based models for these PZT samples are built using PZFlex (Weidlinger Associates Inc, Los Altos, CA) (See Sec. IV) to study the effect of sub-dicing, and to design the size of the sub-diced small unit.

\subsubsection{Second functional model}

This version results from the experiences with the first functional model and the test models above, and attempts to minimize interference from lateral resonances. For this, the first functional model is sub-diced into 0.9 mm x 0.9 mm units using a 100 $\mu$m dicing saw. Only the front matching layer and active layer are sub-diced. The 25 units are acoustically isolated by air kerfs, but electrically grouped by two electrodes (Figure~\ref{fig:singleelement}(d)). The dimensions of each layer are listed in Table~\ref{table:firstmodel}.

\subsubsection{Final model}

With the second functional model above found suitable to minimize the presence of lateral resonance in the frequency range of interest, we studied the variation of passive layers dimensions on the frequency response using 3D FEM models. To increase the bandwidth of the detector, the thicknesses of the front and back matching layer need to be optimized. Three-dimensional FEM models are simulated to estimate the pulse-echo signals of the detectors, from which the frequency responses of the detectors are calculated to arrive at an optimized implementation. We manufactured a final model according to the geometrical parameters achieved as shown in Figure~\ref{fig:singleelement}(b,d). The dimensions of each layer are listed in Table~\ref{table:firstmodel}.

\section{Numerical and experimental methods}

\subsection{Simulation methods used}

\subsubsection{1D KLM model}

An equivalent circuit based 1D KLM model~\cite{Leedom1971} is used to obtain a rough assessment of the required detector performances such as center frequency and bandwidth. 

\subsubsection{3D FEM model}

For a more accurate estimation, coupled partial differential equations for piezoelectricity and acoustic wave propagation through passive material layers of the detector and coupling medium require to be solved~\cite{Qi2000}. This can be appropriately done by using 3D FEM models. In this study, 3D FEM models are built using PZFlex (version 3.0, Weidlinger Associates Inc, Los Altos, CA) as shown in Figure~\ref{fig:singleelement}(a) and (b). The finite element size is defined as 1/30 of the ultrasound wavelength at the designed transducer center frequency (1 MHz). The roughly estimated results from KLM models are used as the starting point for the subsequent iterations between 3D simulations of detector performance and experimental characterization of functional models. The material properties used for FEM simulation are listed in Table~\ref{table:material}.

\subsection{Detector characterization methods}

\subsubsection{Electrical impedance}

In order to ascertain the resonance characteristics, and to study the behaviour of the detector in an electrical measurement chain, the complex electrical impedance is measured using an impedance analyzer (4194A/B, Hewlett-Packard, Palo Alto, California).

\subsubsection{Acoustic frequency response}

Two methods were used to measure the frequency response: (1) a transmit mode using a hydrophone, and (2) a pulse-echo mode. 

In method 1 (Figure~\ref{fig:echo}(a)), the detector immersed in a demineralized water bath is driven by a broadband ultrasonic pulser/receiver (Panametrics 5077PR), and the generated pressure is probed in the far-field using a calibrated broadband needle hydrophone (0.2 mm diameter, bandwidth upto 10 MHz, Precision Acoustic Ltd., Dorchester, UK) with a known receiving transfer function $H_{0}(f)$ .  The frequency response of the detector ($H(f)$) is then calculated by:

\begin{equation}\label{eq4}
H(f) = \displaystyle \frac{\mathrm{FFT}\left\{P(t)\right\}}{H_{0}(f)}
\end{equation}

Method 2 (Figure~\ref{fig:echo} (b)) uses the detector in pulse-echo mode, driven by the pulser-receiver. A stainless steel plate is placed in the far-field of the detector as an acoustic reflector, and the reflected signal (pulse-echo) is measured by the detector. According to the principle of reciprocity for a piezoelectric detector~\cite{Callerama1979, Carstensen1947}, the frequency response of an ultrasound transducer is the same when used as receiver or transmitter. Thus the frequency response of the detector is calculated by the square root of the FFT of the measured pulse-echo signal.

\subsubsection{Directivity}

The directivity of the detector is measured in transmit mode. The detector is mounted in a demineralized water bath and driven by the broadband pulser/receiver. The calibrated needle hydrophone is rotated in the far-field (60 mm) centered on the detector element to probe the emitted ultrasound field through 180$^{\texttt{o}}$. For each scanning position, the peak-to-peak value is determined from the ultrasound pulse recorded by the hydrophone and plotted as a function of scanning angle, giving the directivity of the transducer at its center frequency.

\subsubsection{Sensitivity and minimum detectable pressure}

We measured these parameters using a modified substitution method~\cite{Piras2010} as outlined. We insonify the detector element using an ultrasound transmitter. The pressure incident on the detector element is progressively reduced, by reducing the voltage input to the transmitter. At each incident pressure, the voltage output of the detector is noted while averaging multiple times. This is continued till the lowest input possible. This detector output is defined as the signal. In a next step we note the noise voltage on the detector element without averaging and with no pressure incident on the element. The signal is plotted against pressure input and the trend is extended to intersect the voltage noise floor. This intersection point signifies SNR=1, and the pressure at which this signal is expected/obtained is the minimum detectable pressure.

In the experiments we used a 1 MHz unfocused broadband transducer (V303, Panametrics) as a transmitter. Signals were processed by a prototype low noise pre-amplifier based on Analog Devices ADA4896-2 as the analog front-end and acquired using a high-speed digitizer NI-5752 (National Instruments). A calibrated needle hydrophone was used to ascertain the transfer function of the transmitter and convert the input voltage of the transmitter to known pressure that insonified the detector element.

\subsection{Imaging quality simulation}

To study the resolution and visibility of objects with a photoacoustic tomography (PAT) system employing the optimized detector, and comparing performances using detectors described in the literature, numerical simulations are performed using k-wave Matlab Toolbox~\cite{Cox2005}. 

For the forward simulation, a 2D initial pressure distribution map (1024 x 1024 grid, 20 cm  x 20 cm size) is assigned in a 2D tomographic configuration.  Three disc-shaped objects with diameters of 10 mm, 2 mm and 0.5 mm are located in the center region of the map. Homogenous initial pressure (value 1) is assigned to the objects, a pressure value of zero is given to the rest of the map. Homogenous acoustic properties are assigned to the medium (speed of sound: 1500 m/s, acoustic attenuation: 0, density: 1000 kg/m$^3$.) Those pressures propagating outward are detected by a 5 mm detector rotating around the objects with radius of 10 cm covering 360$^o$ with step size of 2$^o$. The time-domain photoacoustic signals are averaged over the surface of the detector at each detection position and saved for further processing.  

Three detectors are simulated, from: 

\noindent (1) The final model of this work.

\noindent (2) The Kruger group (1 MHz center frequency and 100$\%$ bandwidth)~\cite{Kruger1999}.

\noindent (3) The Oraevsky group (1.25 MHz center frequency and a bandwidth approaching 200$\%$, derived from~\cite{Andreev2003, Ermilov2009}).

For (1), signals received by the detector are convolved with measured impulse response of the final model. For (2) and (3), the received signals are filtered using Gaussian bandpass filters with the corresponding center frequency and bandwidth. Correspondingly, three images are reconstructed by time-reversal of the processed signals.

\section{Results}

\subsection{First functional model performance}

Figure~\ref{fig:firstmodel}(a) shows the measured electrical impedance of the first functional model.  A fundamental thickness resonance at 1.2 MHz together with a series of harmonic thickness resonances at higher frequencies, and a strong radial resonance at 330 kHz together with a second harmonic radial resonance at around 700 kHz, can be observed. The fundamental thickness resonance corresponds to the designed thickness of the active layer (1.625 mm) and the low frequency radial resonance arises due to the large lateral dimensions of the active layer (5 mm x 5 mm)~\cite{Reynolds2003}.

Figure~\ref{fig:firstmodel}(b) shows the measured frequency response of the first functional model using method 1 described above. Two peaks can be observed in the frequency domain response: a 1.2 MHz peak caused by the thickness resonance, and a peak at 330 KHz caused by the radial resonance. Both peaks match with the measured electrical impedance peaks (Figure~\ref{fig:firstmodel}(a)). The first functional model has a center frequency of 1.2 MHz (maximum peak), with a -6 dB bandwidth of 0.8 MHz. The fractional bandwidth is 67$\%$. Due to the radial resonance, a secondary pulse (reverberation-like signal) after the main pulse is visible in the time domain signal, which is manifested as a passband ripple in the frequency domain.

The measured strong lateral resonance of the first functional model is not desirable. First the detector sensitivity is reduced by the lateral mode resonance~\cite{Qi2000}. Second, strong lateral mode causes additional reverberation-like signals (Figure~\ref{fig:firstmodel}(b)), which will adversely influence the image quality of the system~\cite{Yin2010}.  

\subsection{Test models performances}

Figure~\ref{fig:subdicing} shows the measured and simulated electrical impedance of the bare piezoceramic samples with thickness of 1.625 mm and lateral dimensions ranging from 5 mm x 5 mm down to 0.5 mm x 0.5 mm. In general, FEM simulations exhibit good agreement with measurements (due to the manufacturing and measurement difficulties for the smaller bare ceramics, electrical impedance measurements for the 0.5 mm x 0.5 mm ceramic is not available). The fundamental thickness resonance peak for all samples is always located around 1 MHz, while the lateral resonance frequency peaks move towards higher frequencies as the lateral dimensions decrease. For lateral dimensions of 1 mm x 1 mm the radial resonance mode is suppressed. Further reduction in lateral dimensions to 0.5 mm x 0.5 mm does not significantly change the amplitude and location of the thickness resonance peak (Figure~\ref{fig:subdicing}(e) and (f)). Since more active material is lost with finer subdicing, leading to detection sensitivity loss, we prefer 1 mm x 1 mm as the final choice. 

\subsection{Second functional model}

The measured electrical impedance of the second functional model in water is compared with 3D FEM simulation in Figure~\ref{fig:firstdice}(a). Both simulation and measurement show that the lateral resonance is not visible in the electrical impedance curves as expected. The fundamental thickness resonance is located at around 1 MHz. 

The frequency response of the detector measured using method 2 compared with 3D FEM simulations is shown in Figure~\ref{fig:firstdice}(c). Both measurement and simulation shows the detector has a center frequency of 0.9 MHz, however with a -6 dB fractional bandwidth of 48$\%$, which is low for photoacoustic applications. 

\subsection{Final model}

The starting point for the thickness optimization is based on the 1-D KLM model~\cite{Leedom1971}; this gives the optimized front matching layer thickness ($t_{ML-F}$) of 0.58 mm and back matching layer thickness ($t_{ML-B}$) of 0.54 mm as used for the first functional model (Figure~\ref{fig:firstmodel}(b)). In the first phase of the optimization, $t_{ML-B}$ is kept constant (0.54 mm), and $t_{ML-F}$ is tuned to obtain the largest bandwidth (Figure~\ref{fig:optimization}(a)-(c)). The optimum $t_{ML-F}$ value is then determined (0.70 mm). The second phase is to keep the optimized $t_{ML-F}$ constant, and to optimized $t_{ML-B}$. After the two phase optimization, the optimum $t_{ML-F}$ and $t_{ML-B}$ are determined (Figure~\ref{fig:optimization}(d)). 

The simulation results in Figure~\ref{fig:optimization} show that the bandwidth of the detector is increased from around 50$\%$ (Figure~\ref{fig:optimization}(a)) to around 70$\%$ (Figure~\ref{fig:optimization}(c)) by optimizing the front matching layer. Further optimization of the back matching layer increases the bandwidth of the detector to more than 80$\%$ (Figure~\ref{fig:optimization}(d)), which is generally high for a PZT detector with single front matching layer~\cite{Matthaei1980}.

\subsubsection{Frequency response}

The measured frequency response and time domain signal in the farfield (60 mm away) show good agreement with simulation results in Figures~\ref{fig:finalmodel}(a) and~\ref{fig:finalmodel}(b). Compared to the ultrasound pulse for the first functional model (Figure~\ref{fig:firstmodel}(b)), the optimized pulse is considerably shorter in length. The measured center frequency is 1 MHz, which is slightly higher than the model predicted 0.9 MHz (Figure~\ref{fig:finalmodel}(b)). The measured -6 dB bandwidth ranges from 0.4 MHz to 1.25 MHz, which is also slightly larger than the simulation result (0.55 MHz - 1.25 MHz), while the fractional bandwidths are similar for both (80$\%$). 
   
\subsubsection{Directivity}

Figure~\ref{fig:finalmodel}(c) shows the measured directivity of the final model. The measurement results compare well with simulations, both showing a -6 dB acceptance angle around 20$^{\texttt{o}}$ at 1 MHz. When a 5 mm diameter hemispherical acoustic lens is placed on top of the detector, the directivity angle of the detector is expected to be enlarged to around 60$^{\texttt{o}}$~\cite{Xia2011}.

\subsubsection{Sensitivity}

The end-of-cable minimum detectable pressure of the detector-electronics was estimated to be 0.5 Pa. As seen the Figure~\ref{fig:finalmodel}(d)) the signal trend is extrapolated to intersect the noise floor to provide the point at which SNR=1. The corresponding pressure can be read off on the x-axis. 

\subsection{Imaging quality}

The reconstructed images from the three cases are shown: for our detector in Figure~\ref{fig:resolution}(b), Kruger system~\cite{Kruger1999} in Figure~\ref{fig:resolution}(c), and the Oraevsky system~\cite{Andreev2003, Ermilov2009} in Figure~\ref{fig:resolution}(d). It is clearly shown that Oraevsky detector, due to the superior bandwidth of the detector, possesses excellent image quality and faithfully recovers the original initial pressure distribution. Ours and Kruger's system faithfully recover the 2 mm object. The 10 mm object is visualized with edge enhancement. The resolution of three systems is around 1-2 mm, which can be estimated from the reconstruction of the sub-resolution 0.5 mm object (see profiles in Figure~\ref{fig:resolution}(e)).

\section{Discussion}

The final optimized single-element detector has a fractional bandwidth of 80$\%$, which matches with the model prediction. The center frequency of the detector is around 1 MHz and -6 dB bandwidth is from 0.40 MHz to 1.25 MHz. From simulation of system resolution and image quality (Figure~\ref{fig:resolution}(b)), it is found that the object with diameter of 2 mm can be faithfully recovered, and the 10 mm object can be visualized with acceptable distortions (edge enhancement and ring-shaped artifacts). The resolution of our system reaches the designed goal of 1-2 mm (Figure~\ref{fig:resolution}(e)). The simulations are performed with a high optical contrast of the object, which gives large signal-to-noise ratios. This condition is applicable for a sensitive detector like ours: frequencies beyond the -6 dB bandwidth of the detector will still contribute to signal detection, which makes the resolution of our system better than the value calculated from Eq.(2)~\cite{Andreev2003} taking only -6 dB bandwidth of the detector into account. The resolution metrics used via k-wave simulations do not necessarily line up with those from Eq.(2) because Eq.(2) is derived only from the frequency contents of the photoacoustic signals generated from objects with varying sizes. This approach gives an indication for designing the ultrasound detectors, while imaging approach via k-wave simulation is suitable to characterize the overall system resolution. This indicates that the present detector is suitable to detect tumors during the beginning of the vascular phase. The image quality can be further improved by enlarging the bandwidth of the detector using two or more front matching layers~\cite{Matthaei1980}. However, a larger bandwidth reduces the sensitivity of the detector, and increases complexity and expense during transducer design and development and in  final production.

The directivity angle of the transducer is around 20$^{\texttt{o}}$, which can be increased to around 60$^{\texttt{o}}$ using an acoustic lens. For simplicity, considering the breast as a hemisphere with diameter of 10 cm, the transducer with directivity angle of 60$^{\texttt{o}}$ is required to be placed only 5 cm away from the breast. This distance is suitable for the application.

We have consolidated the most important output characteristics of the optimized ultrasound detector in Table~\ref{table:detectors}, in row 7 described as the PAM-II system. We have also provided specifications of various ultrasound detectors described in literature for comparison. The MDP value of our detector is 0.5 Pa, which is 160 times lower than for the detector used in the first generation of the Twente Photoacoustic Mammoscope previously developed in our group~\cite{Manohar2005}. To the best of our knowledge, this is the lowest measured MDP reported for a detector in photoacoustic breast imaging (Table~\ref{table:detectors}). The total noise performance of the system could possibly further be improved with better electrical shielding and grounding.

The bandwidth of our detector is 80$\%$, which is slightly broader than the detectors used by Kruger group and Kyoto University in their breast imagers (See Table~\ref{table:detectors}). The detector from the Oraevsky group has an ultrabroad bandwidth approaching 200$\%$ (estimated from Refs.~\cite{Andreev2003, Ermilov2009}), which provides impressive image quality as shown in Figure~\ref{fig:resolution}(d). The image quality of our detector can be improved with the use of a deconvolution operation as is performed by several groups~\cite{YWang2004,KWang2012, Rosenthal2011, Jose2012} to compensate for finite bandwidths effects. Further breast tumors are known to be heterogeneous with a scattered distribution of absorbing regions. In such a case our detector is eminently suited to faithfully image a collection of small absorbing structures, which makes up the tumor mass. We are aware that there is room for improvement, nevertheless our design strategy has resulted in a detector with acceptable bandwidth which is well suited for sensitive clinical breast imaging due to its unprecedented high sensitivity. 

Detector arrays will be manufactured based on the single-element detector described in this work for use in the second version of the Twente Photoacoustic Mammoscope (PAM-II)~\cite{Piras2010}.

\section{Conclusion}

A single-element PZT, large-aperture, sensitive and broadband detector is designed and developed for photoacoustic tomography of the breast. Finite-element based models are used to optimize an initial detector to reduce the radial resonance and optimize the bandwidth of the detector. The center frequency and -6 dB fractional bandwidth of the optimized detector are 1 MHz and around 80$\%$, respectively. The minimum detectable pressure is 0.5 Pa, which is more than two orders of magnitude lower than in our first generation photoacoustic breast imaging system and among the lowest reported in the literature. Detector arrays will be manufactured based on the design of this single-element detector for use in the second version of the Twente Photoacoustic Mammoscope.

\section{Acknowledgments}
The financial support of the Agentschap NL Innovation–Oriented Research Programme (IOP) Photonic Devices under the HYMPACT Project (IPD083374); MIRA Institute for Biomedical Technology and Technical Medicine, University of Twente is acknowledged. Franc van den Adel is acknowledged for his useful suggestions and discussions.

\newpage

\section*{List of Figure Captions}

{\sffamily \textbf{Figure 1:} Schematics of a 2D photoacoustic tomography system showing the necessity for the acceptance angle to encompass the object for coherent signals detection from all angular position in performing reconstruction.}

{\sffamily \textbf{Figure 2:} (a) Schematic of first functional model. (b) Schematic of subdiced second functional model and final model.(c) Photograph of the bare PZT samples with different lateral dimensions (label below). Two triangular samples are shown in the photograph, however, no related result is reported in this work. (d) Photograph of the final single-element model.}

{\sffamily \textbf{Figure 3:} Schematics of the setups for detector frequency response measurements. (a) Transmit mode. (b) Pulse-echo mode.}

{\sffamily \textbf{Figure 4:} First functional model performance. (a) Electrical impedance of the first functional model measured with water load. (b) Measured transmission impulse response and frequency transfer function of the first functional model using a hydrophone. (time domain left axis, frequency domain right axis)

{\sffamily \textbf{Figure 5:} Test samples of PZT: measured and simulated electrical impedance in air with lateral dimensions (a) 5 mm x 5 mm; (b) 4 mm x 4 mm; (c) 3 mm x 3 mm; (d) 2 mm x 2 mm; (e) 1 mm x 1 mm and (f) 0.5 mm x 0.5 mm. No measured impedance available for 0.5 mm x 0.5 mm PZT due to the practical limitations in manufacturing and measuring.}

{\sffamily \textbf{Figure 6:} Second functional model performance: (a) Measured and simulated electrical impedance in water. (b) Measured and simulated pulse-echo ultrasound signal. The reflector is placed in the far-field of the detector. The time delay is removed. (c) Measured and simulated frequency response.}

{\sffamily \textbf{Figure 7:} Towards optimized final model: simulated pulse-echo signal and the frequency response of the sub-diced detector with different front- and back- matching layer thicknesses. (a) $t_{ML-F}$: 0.58 mm, $t_{ML-B}$: 0.54 mm. (b) $t_{ML-F}$: 0.55 mm, $t_{ML-B}$: 0.54 mm. (c) $t_{ML-F}$: 0.70 mm, $t_{ML-B}$: 0.54 mm. (d) $t_{ML-F}$: 0.70 mm, $t_{ML-B}$: 0.48 mm. }

{\sffamily \textbf{Figure 8:} Final model performance: (a) Measured and simulated far-field pulse of the final model time-shifted to origin. For the measurement, the pulse is probed using a calibrated broadband needle hydrophone in the far-field at distance 60 mm, on center axis. (b) Measured and simulated frequency response of the detector. (c) Measured and simulated directional sensitivity. (d) Measured sensitivity and minimum detectable pressure (MDP).}

{\sffamily \textbf{Figure 9:} (a) Initial pressure distribution used in the forward simulation. The gray dashed circle indicates the detector scanning positions. (b) Reconstructed image using signals detected by the final model, (c) by the transducer with 1 MHz center frequency and 100$\%$ fractional bandwidth (Kruger et al 1999~\cite{Kruger1999}) and (d) by the transducer with 1.25 MHz center frequency and 200$\%$ fractional bandwidth (Andreev et al 2003~\cite{Andreev2003}and Ermilov et al 2009~\cite{Ermilov2009}). (e)Profiles at position X=0 mm from the initial pressure distribution in (a) and reconstructed images from (b),(c) and (d).}

\newpage
\section*{List of Table Captions}

{\sffamily \textbf{Table 1:} Properties of the materials used for the detector in 3D FEM simulations. Properties of the PZT material are from reference~\cite{Sherrit1997, Neer2010}, and the properties of the matching and backing layers are from reference~\cite{Neer2010}.}

{\sffamily \textbf{Table 2:} Layer thicknesses of functional models (Figure~\ref{fig:singleelement}(a) and (b)), each layer has a 5 mm x 5 mm square-shape surface.}

{\sffamily \textbf{Table 3:} List of US detectors used by different groups in the photoacoustic (thermoacoustic) systems for breast imaging.}

\newpage

\begin{figure}
  \centering
  \includegraphics[angle=0,width=0.6\linewidth]{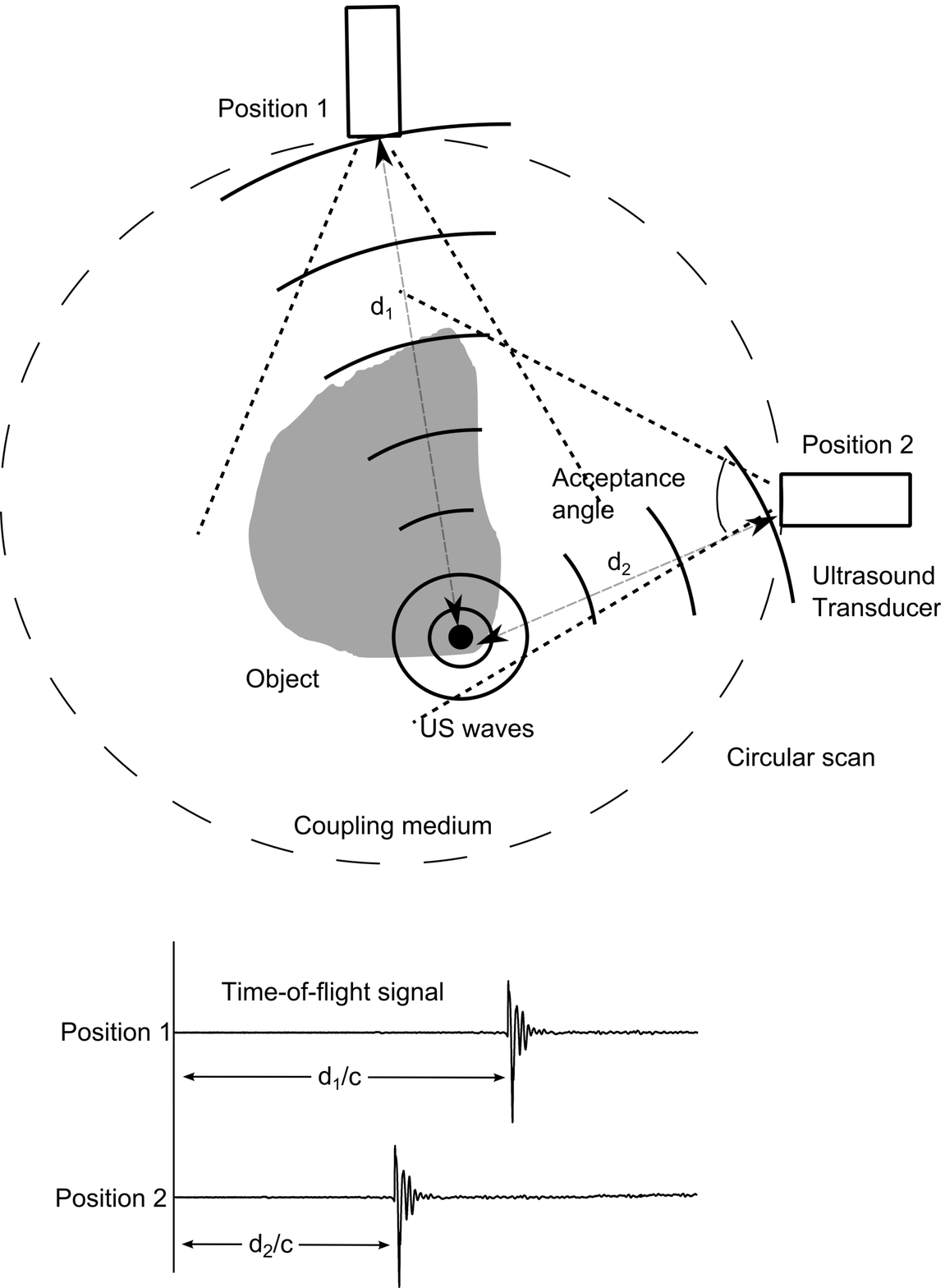}
  \caption{}
  \label{fig:CT}
\end{figure}
\vspace{80mm}
\begin{center}
\end{center}
\newpage

\begin{figure}
  \centering
  \includegraphics[angle=0,width=1\linewidth]{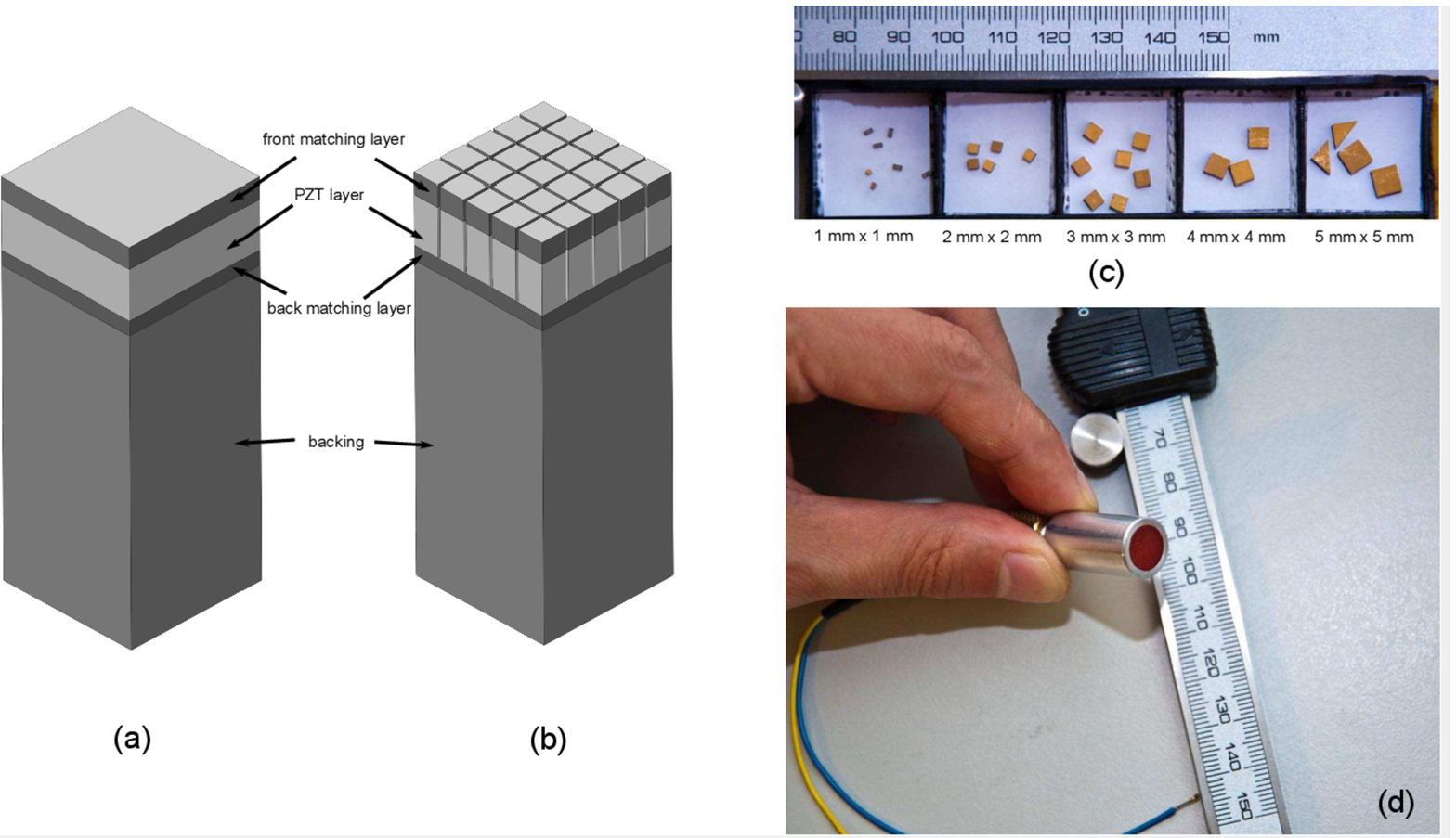}
  \caption{}
  \label{fig:singleelement}
\end{figure}
\vspace{80mm}
\begin{center}
\end{center}
\newpage

\begin{figure}
  \centering
  \includegraphics[angle=0,width=1\linewidth]{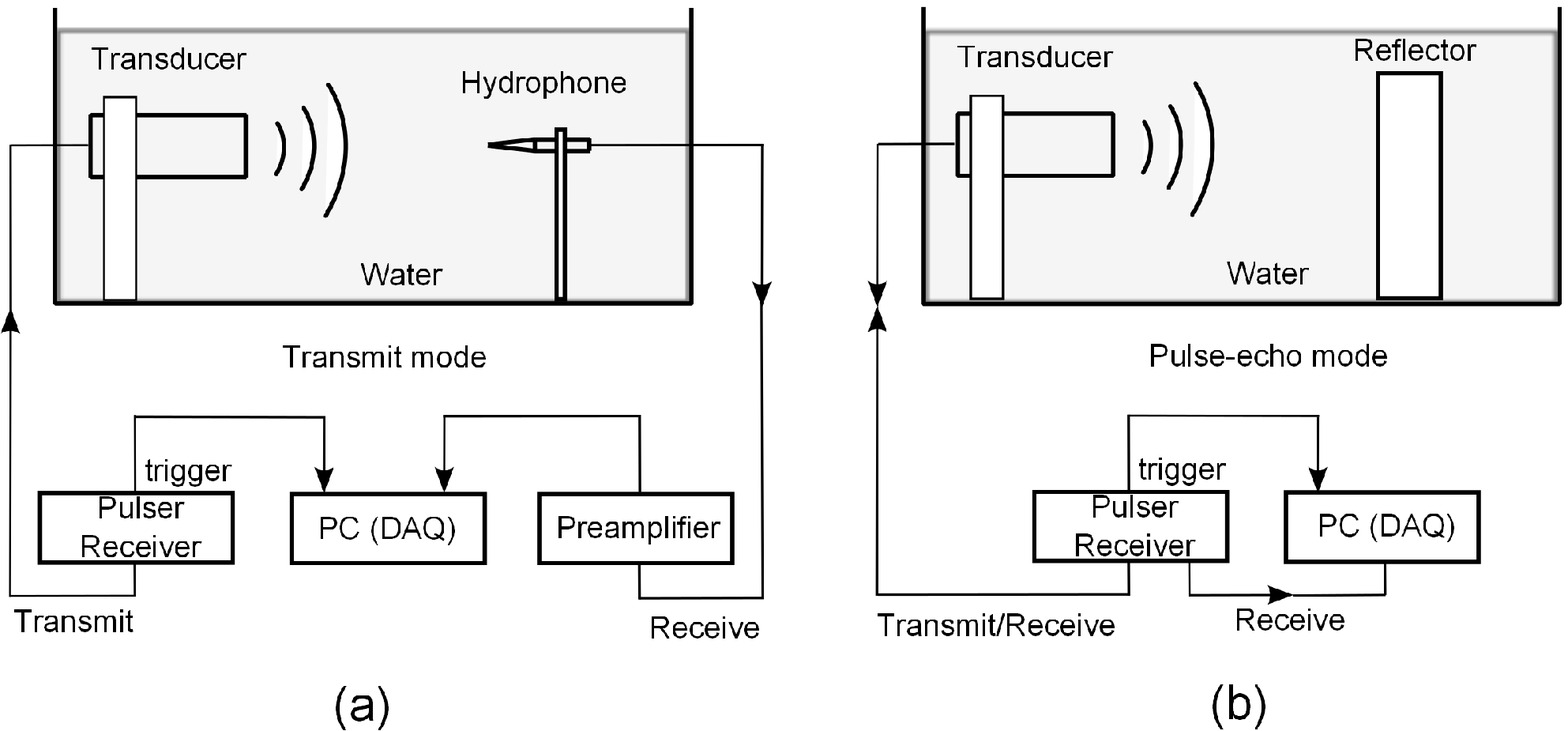}
  \caption{}
  \label{fig:echo}
\end{figure}
\vspace{80mm}
\begin{center}
\end{center}
\newpage

\begin{figure}[!ht]
  \centering
  \includegraphics[angle=0,width=1\linewidth]{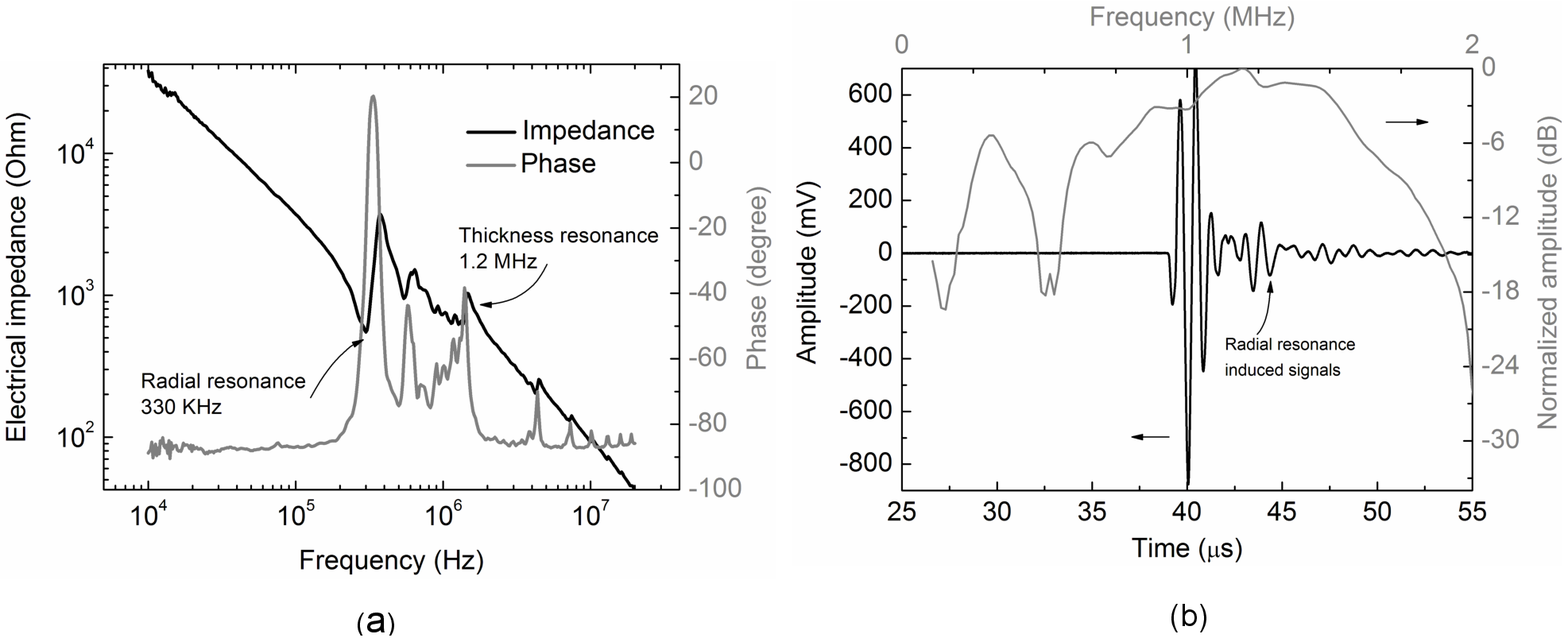}
  \caption{}
  \label{fig:firstmodel}
\end{figure}
\vspace{80mm}
\begin{center}
\end{center}
\newpage

\begin{figure}
  \centering
  \includegraphics[angle=180,width=1.0\linewidth, angle = 180]{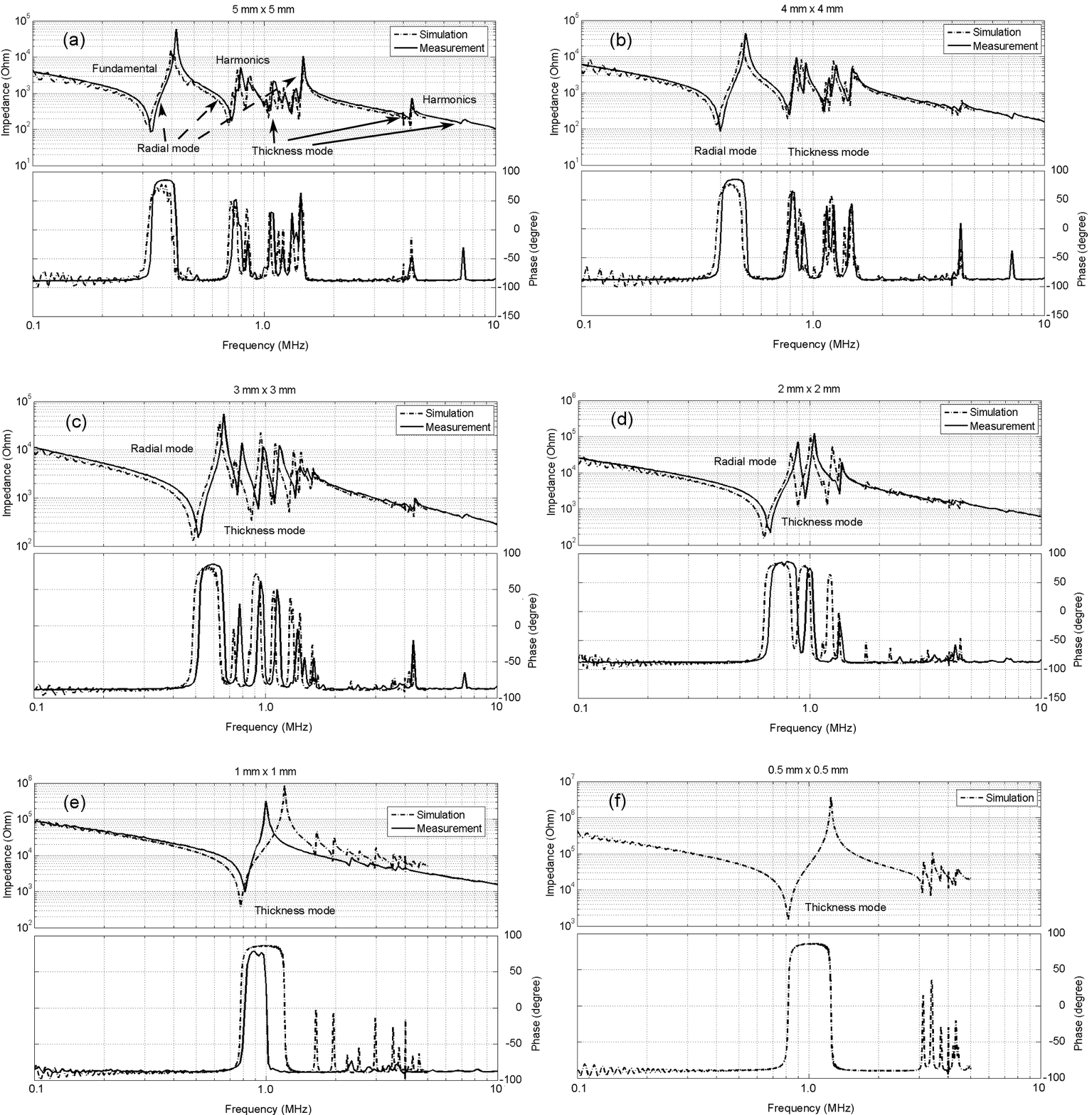}
  \caption{}
  \label{fig:subdicing}
\end{figure} 
\vspace{80mm}
\begin{center}
\end{center}
\newpage

\begin{figure}
  \centering
  \includegraphics[angle=0,width=0.8\linewidth]{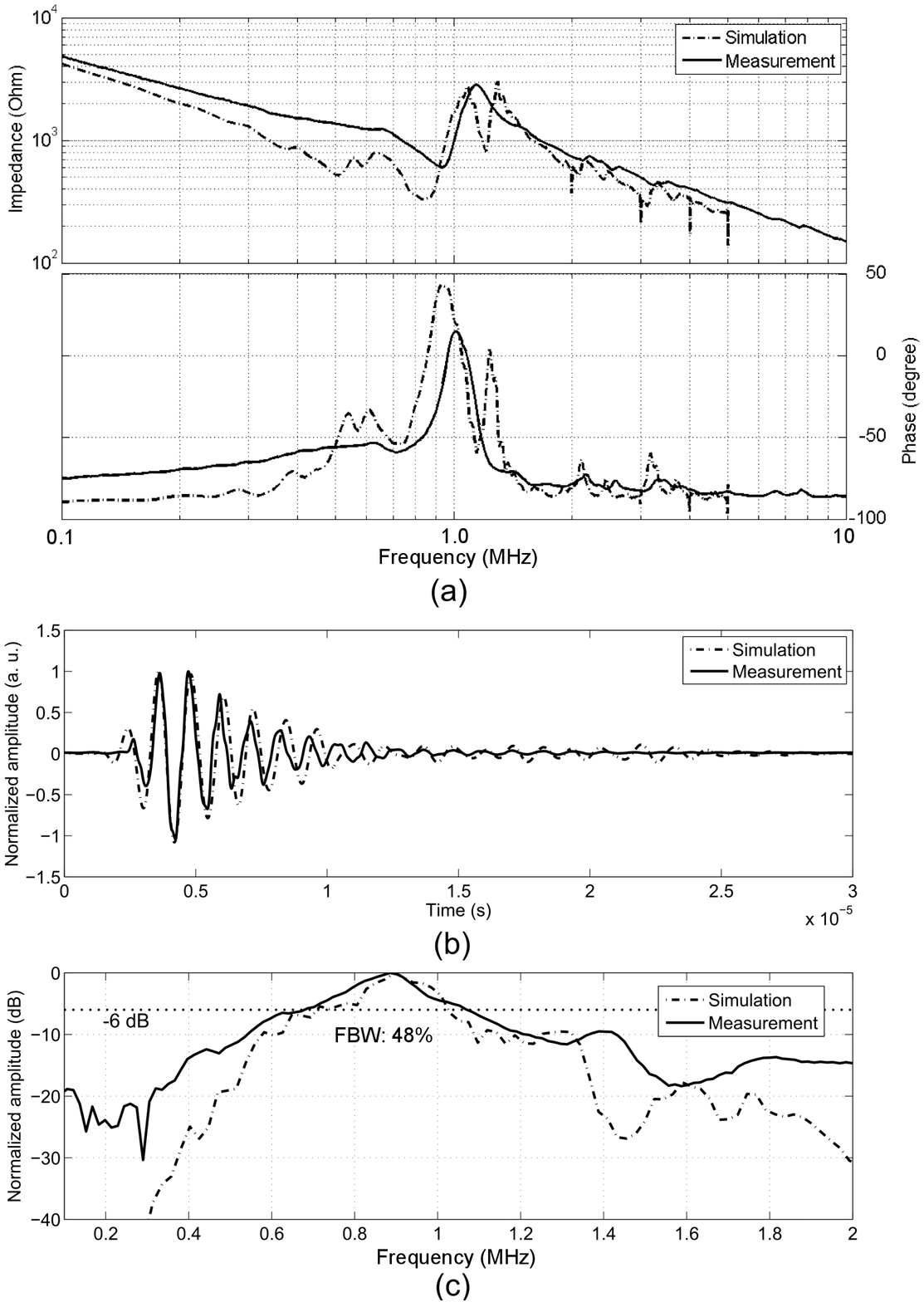}
  \caption{}
  \label{fig:firstdice}
\end{figure}
\vspace{80mm}
\begin{center}
\end{center}
\newpage

\begin{figure}
  \centering
  \includegraphics[angle=0,width=1\linewidth]{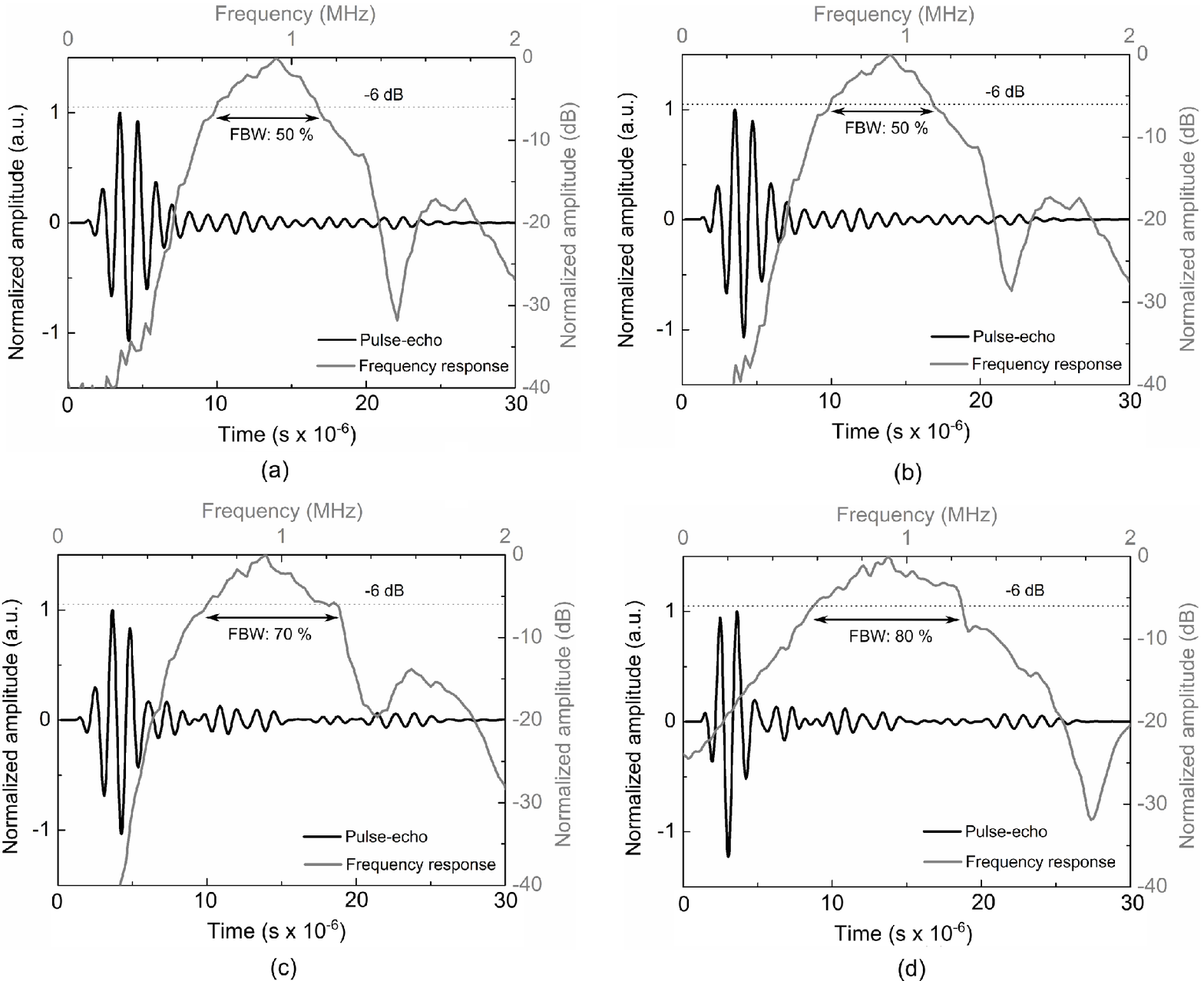}
  \caption{}
  \label{fig:optimization}
\end{figure}
\vspace{80mm}
\begin{center}
\end{center}
\newpage

\begin{figure}
  \centering
  \includegraphics[angle=0,width=1\linewidth]{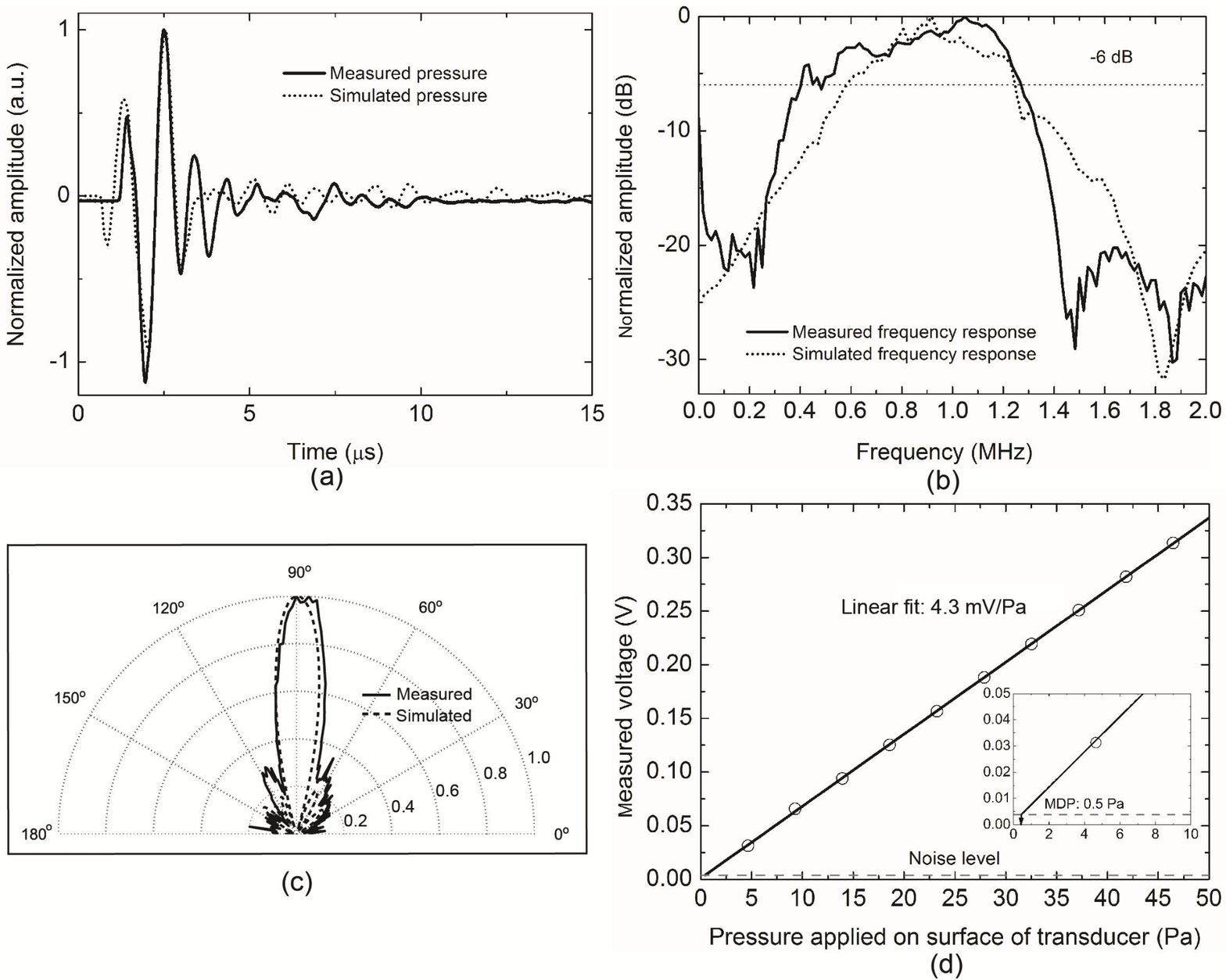}
  \caption{}
  \label{fig:finalmodel}
\end{figure}
\vspace{80mm}
\begin{center}
\end{center}
\newpage

\begin{figure}
  \centering
  \includegraphics[angle=0,width=1\linewidth]{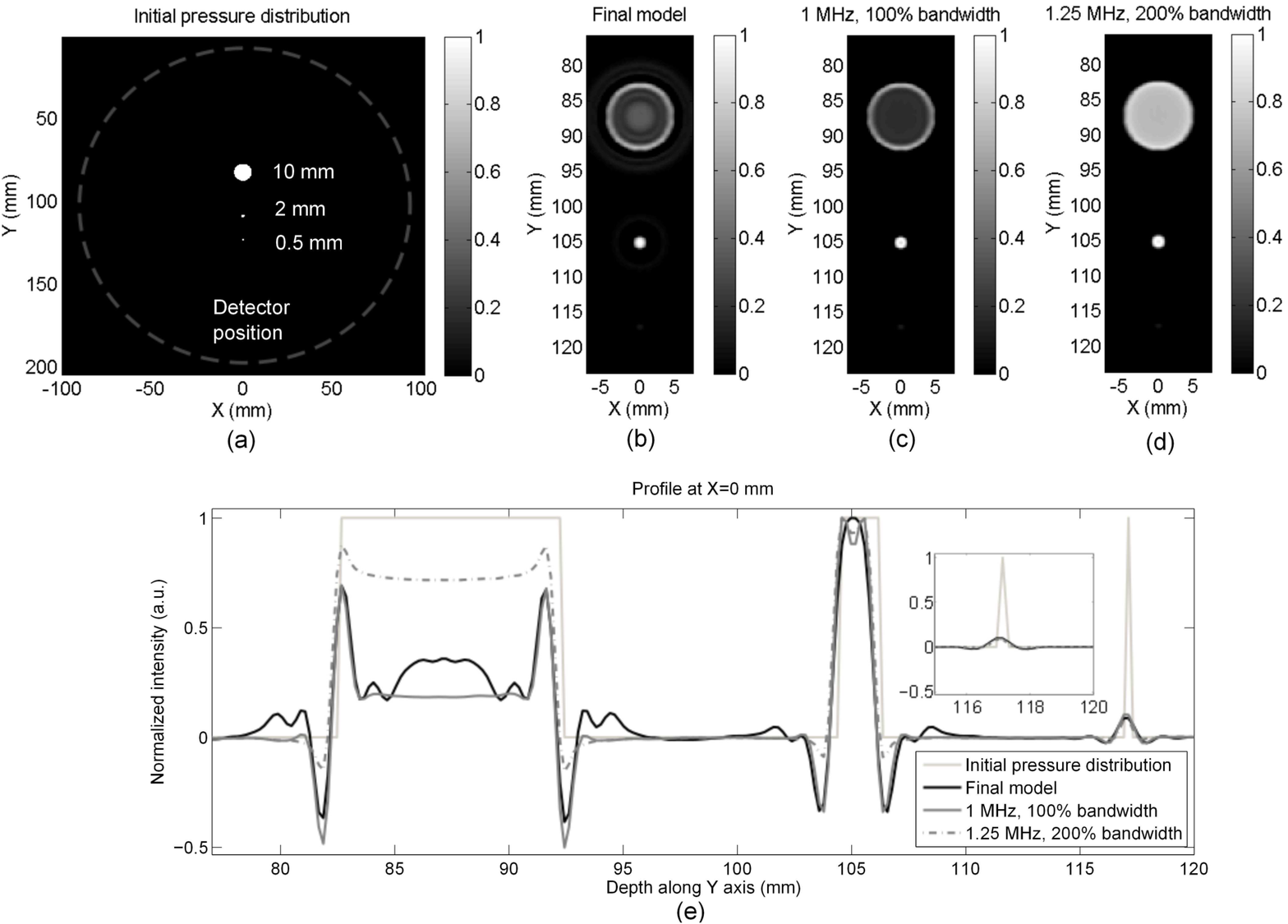}
  \caption{}
  \label{fig:resolution}
\end{figure}
\vspace{80mm}
\begin{center}
\end{center}
\newpage

\section*{Tables}
\normalsize
\renewcommand{\baselinestretch}{1}

\begin{table*}[ht!]
\centering 
\caption{}
\label{table:material}
\begin{tabular}{@{}llll}\hline\hline
Properties & Matching layers & Active layer & Backing \\
\hline
Material  &  Electrical conductive epoxy &  PZT(CTS 3203HD) & Elastosil \\
Impedance (MRayl)&  6.5 & 38 & 3.4 \\
Density (kg~m$^{-3}$) & 3140  & 7800 & 1870 \\
Longitudinal velocity (m~s$^{-1}$) & 2068  & -  & 1818 \\
Shear velocity (m~s$^{-1}$) & 994 & - & 873 \\
Elastic compliance  & - & $s^{E}_{11}$=1.56, $s^{E}_{12}$=-0.420, & - \\
 (m$^{2}$/Nx10$^{-11}$) &  & $s^{E}_{13}$=-0.823, $s^{E}_{33}$=1.89, &  \\ 
 &  & $s^{E}_{33}$=3.92, $s^{E}_{66}$=3.98 &  \\ 
Piezoelectric strain coefficient  & - & $d_{13}$=2.95, $d_{33}$=5.64, & - \\
 (m$^{2}$/Nx10$^{-10}$) &  & $d_{15}$=5.60 &  \\
Relative permitivity & - & $K^{T}_{11}$=2417, $K^{T}_{33}$=3331 & - \\ 
Dielectric loss & - & 0.028 & - \\ 
Mechanical quality factor & - & 66 & - \\ 
 
\hline\hline
\end{tabular}
\end{table*}
\newpage

\begin{table*}[ht!]
\centering 
\caption{}
\label{table:firstmodel}
\begin{tabular}{@{}llll}\hline\hline
Layer description & Material & $1^{\mathrm{st}}$ and $2^{\mathrm{nd}}$ functional models & Final model \\
 &  & thickness (mm) & thickness (mm) \\
\hline
Front matching layer & Electrical conductive epoxy & 0.590 & 0.700 \\ 
Active layer & CTS 3203HD & 1.625 &1.625  \\
Back matching layer & Electrical conductive epoxy & 0.788 & 0.480  \\
Backing & Elastosil & 10 & 10 \\

\hline\hline
\end{tabular}
\end{table*}
\newpage

\begin{table}[ht!]
\centering 
\caption{}
\label{table:detectors}
\begin{tabular}{@{}llllllll}\hline\hline
Groups & System & Detector & Element & Active & Center& BW & MDP \\
 &  & elements &  geometry & material & frequency  & & (Pa)  \\
  &  &  &  (mm)  &  & (MHz)  & & \\
\hline

(1) Kruger group & TA~\cite{Kruger2000} & 64 & $\o$~13  & 1-3 composite PZT$^{\rm \ast }$ & 1  & 70$\%$$^{\rm \ast }$ & - \\ 
(2) Kruger group & PA~\cite{Kruger2010} & 128  & $\o$~3  & 1-3 composite PZT$^{\rm \ast }$ & 5  & 70$\%$$^{\rm \ast }$ & - \\
(3) Wang group & TA$\&$PA~\cite{Pramanik2008} & 1  & $\o$~13/6 & Piezocomposite$^{\rm \dagger}$ & 2.25  & 60-120$\%$$^{\rm \dagger}$ & - \\
(4) Oraevsky group & LOIS-64~\cite{Ermilov2009} & 64  & 20 x 3 & PVDF & 1.25  & 170$\%$ $^{\rm \S}$ & $\rm \ast\ast$ \\
(5) Kyoto University & PAM~\cite{Fukutani2011} & 345  & 2 x 2 & Piezocomposite & 1  & $\geq$70$\%$ & - \\
(6) University of Twente & PAM~\cite{Manohar2004} & 590  & 2 x 2 & PVDF & 1  & 130$\%$ & 80\\
(7) University of Twente & PAM-II$^{\rm \ddagger }$& 1  & 5 x 5 & CTS 3203HD & 1  & 80$\%$ & 0.5\\

\hline\hline
\end{tabular}

\begin{flushleft}
 $^{\rm \ast }$ R. Kruger (2012), private communication.\\ 
 $^{\rm \dagger}$ L. V. Wang and M. Pramanik (2012), private communication.\\
 $^{\rm \S}$ A. A. Oraevsky (2012), private communication.\\
 $^{\rm \ast\ast}$ Various values are reported in literature including measured~\cite{Lamela2011}, estimated~\cite{Andreev2003, Andreev2000}, and a measured MDP of 1.8 Pa from A. A. Oraevsky (2012), private communication.\\
 $^{\rm \ddagger }$ Planned.\\
\end{flushleft}
 
\end{table}
\newpage

\end{document}